# Two-dimensional $O(3)$ $\sigma$-model up to correlation length $10^5$

Sergio Caracciolo[a*], Robert G. Edwards[b], Andrea Pelissetto[c] and Alan D. Sokal[d]

[a]Dipartimento di Fisica and INFN, Università degli Studi di Lecce, Lecce 73100, ITALIA

[b]SCRI, Florida State University, Tallahassee, FL 32306, USA

[c]Dipartimento di Fisica and INFN – Sezione di Pisa, Università degli Studi di Pisa, Pisa 56100, ITALIA

[d]Department of Physics, New York University, 4 Washington Place, New York, NY 10003, USA

We carry out a high-precision Monte Carlo simulation of the two-dimensional $O(3)$-invariant $\sigma$-model at correlation lengths $\xi$ up to $\sim 10^5$. Our work employs a new and powerful method for extrapolating finite-volume Monte Carlo data to infinite volume, based on finite-size-scaling theory. We compare the extrapolated data to the renormalization-group predictions. The deviation from asymptotic scaling, which is $\approx 25\%$ at $\xi \sim 10^2$, decreases to $\approx 4\%$ at $\xi \sim 10^5$.

We study the lattice $\sigma$-model taking values in the unit sphere $S^{N-1} \subset \mathbb{R}^N$, with nearest-neighbor action $\mathcal{H}(\boldsymbol{\sigma}) = -\beta \sum \boldsymbol{\sigma}_x \cdot \boldsymbol{\sigma}_y$. Perturbative renormalization-group computations through three loops [1,2] predict that the exponential correlation length (inverse mass gap) $\xi^{(exp)}$, the second-moment correlation length $\xi^{(2)}$, and the susceptibility $\chi$ behave for $N = 3$ (in infinite volume) as

$$\xi^{\#}(\beta) \;=\; C_{\xi^{\#}} \frac{e^{2\pi\beta}}{2\pi\beta} \left[1 - \frac{0.091}{\beta} + \cdots\right] \qquad (1)$$

$$\chi(\beta) \;=\; C_{\chi} \frac{e^{4\pi\beta}}{(2\pi\beta)^4} \left[1 - \frac{0.001}{\beta} + \cdots\right] \qquad (2)$$

as $\beta \to \infty$. The nonperturbative constant $C_{\xi^{(exp)}}$ has been computed recently using the thermodynamic Bethe Ansatz [3]:

$$C_{\xi^{(exp)}} \;=\; 2^{-5/2} \left(\frac{e^{1-\pi/2}}{8}\right) \;. \qquad (3)$$

The remaining nonperturbative constants are known analytically only at large $N$ [4]:

$$C_{\xi^{(2)}}/C_{\xi^{(exp)}} \;=\; 1 - \frac{0.003225}{N} + O(1/N^2) \qquad (4)$$

$$C_{\chi} \;=\; \frac{\pi}{16}\left[1 - \frac{4.267}{N} + O(1/N^2)\right] \qquad (5)$$

*Speaker at the conference.

A high-precision Monte Carlo study [5] yields $C_{\xi^{(2)}}/C_{\xi^{(exp)}} = 0.9993 \pm 0.0006$ for $N = 3$, in good agreement with (4). Previous studies up to $\xi \sim 100$ agree with these predictions to within about 20–25% [6].

In order to extrapolate finite-volume Monte Carlo data to infinite volume, we used a novel method [7] based on the finite-size-scaling Ansatz

$$\frac{\mathcal{O}(\beta, sL)}{\mathcal{O}(\beta, L)} \;=\; F_{\mathcal{O}}\Big(\xi(\beta, L)/L\,;\, s\Big)\;, \qquad (6)$$

which is correct up to terms of order $\xi^{-\omega}$ and $L^{-\omega}$; here $\mathcal{O}$ is any long-distance observable, $s$ is a fixed scale factor (usually $s = 2$), $L$ is the linear lattice size, $F_{\mathcal{O}}$ is a universal function, and $\omega$ is a correction-to-scaling exponent. For similar extrapolation methods, see [8,9].

Details of our simulation and of the extrapolation process can be found in Sokal's talk at this conference and in [10]. Our preferred fit is shown in Figure 1, where we compare also with the perturbative prediction

$$F_{\xi}(x; s) = s\left[1 - \frac{\log s}{8\pi x^2} - \left(\frac{\log^2 s}{128\pi^2} + \frac{\log s}{16\pi^2}\right)\frac{1}{x^4}\right] \qquad (7)$$

valid for $x \gg 1$.

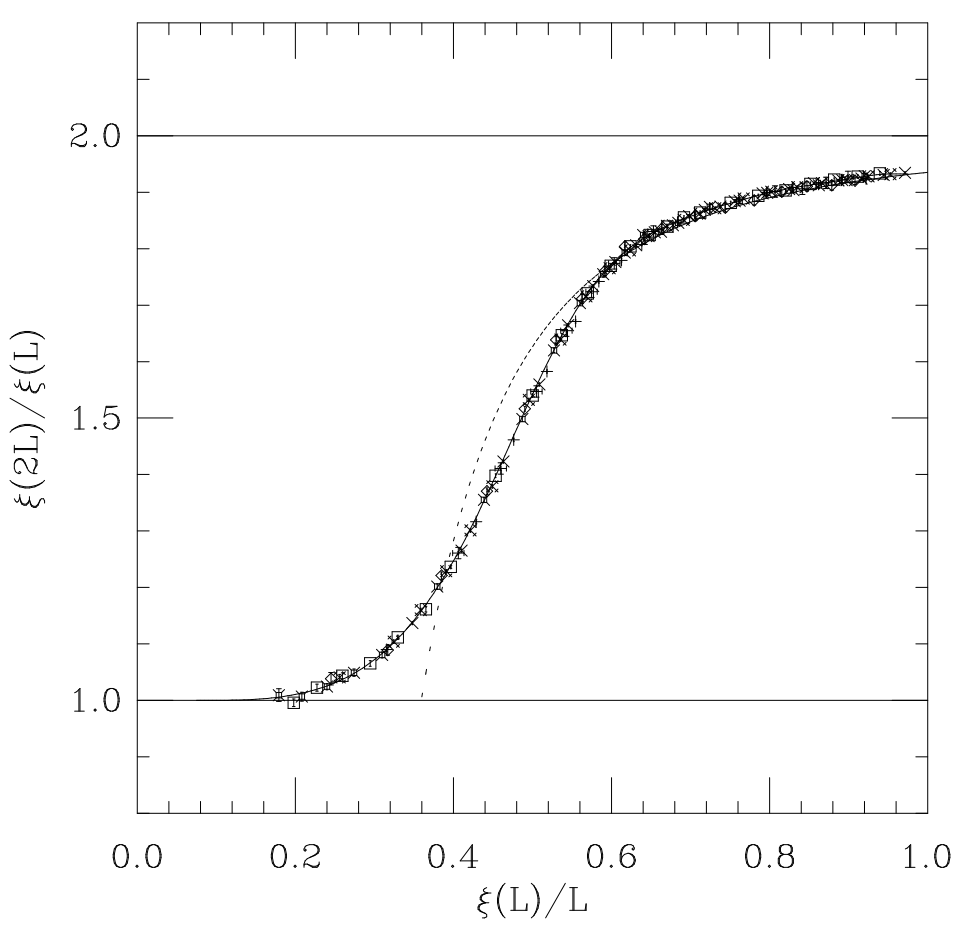

Figure 1. $\xi(\beta,2L)/\xi(\beta,L)$ versus $\xi(\beta,L)/L$. Different symbols indicate different lattice sizes $L$. Dashed curve is the perturbative prediction (7).

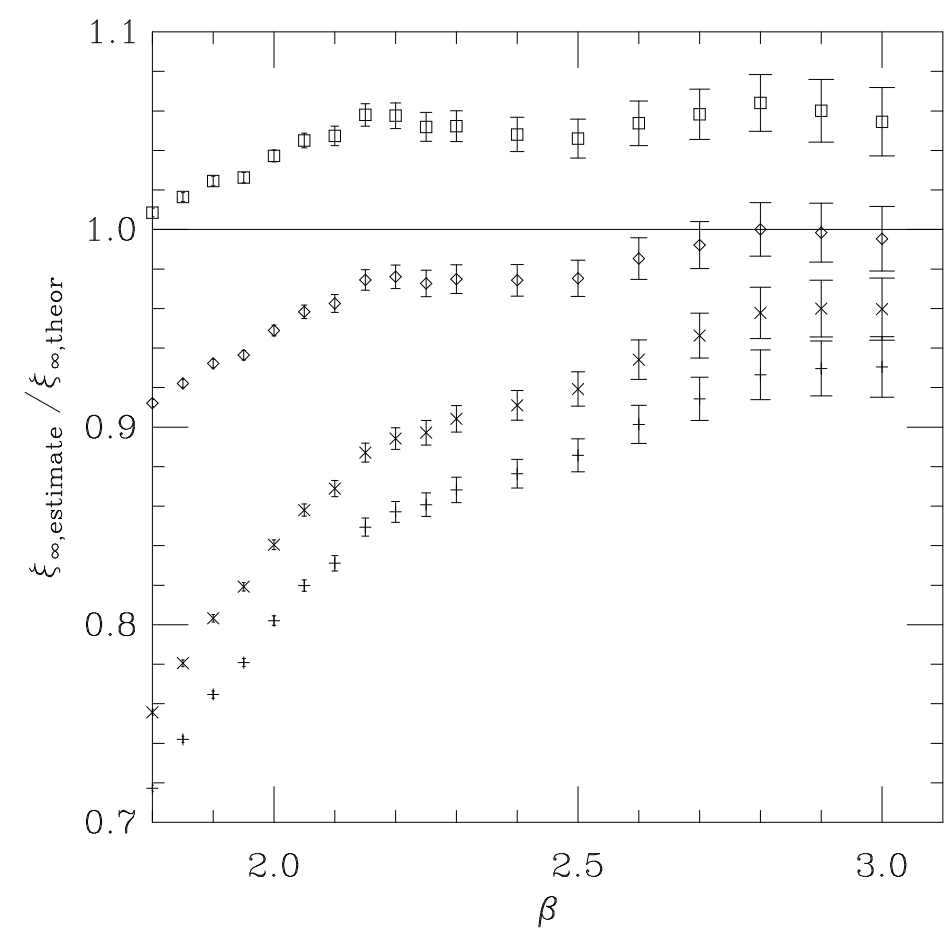

Figure 2. $\xi^{(2)}_{\infty,estimate}/\xi^{(2)}_{\infty,theor}$ versus $\beta$. Error bars are one standard deviation (statistical error only). There are four versions of $\xi^{(2)}_{\infty,theor}$: standard perturbation theory in $1/\beta$ gives points + (2-loop) and × (3-loop); "improved" perturbation theory in $1-E$ gives points □ (2-loop) and ◇ (3-loop).

The extrapolated values $\xi^{(2)}_{\infty}$ from different lattice sizes at the same $\beta$ are consistent within statistical errors: only one of the 24 $\beta$ values has a $\chi^2$ too large at the 5% level; and summing all $\beta$ values we have $\chi^2 = 86.56$ (106 DF, level = 92%).

In Figure 2 (points + and ×) we plot $\xi^{(2)}_{\infty,estimate}$ divided by the two-loop and three-loop predictions (1)–(4). The discrepancy from three-loop asymptotic scaling, which is $\approx 16\%$ at $\beta = 2.0$ ($\xi \approx 200$), decreases to $\approx 4\%$ at $\beta = 3.0$ ($\xi \approx 10^5$). This is roughly consistent with the expected $1/\beta^2$ corrections. Notice that the points in the curve fluctuate less than what one would expect on the basis of the reported error bars; this is due to a strong statistical correlation of the estimates for the points with higher values of $\beta$. Probably also the slight bump at $2.3 \lesssim \beta \lesssim 2.6$ is spurious, arising from correlated statistical or systematic errors.

We can also try an "improved expansion parameter" [11,2,13] based on the energy $E = \langle \boldsymbol{\sigma}_0 \cdot \boldsymbol{\sigma}_1 \rangle$. First we invert the perturbative expansion [12,2]

$$E(\beta) = 1 - \frac{1}{2\beta} - \frac{1}{16\beta^2} - \frac{0.038512}{\beta^3} + O(1/\beta^4) \quad (8)$$

and substitute into (1); this gives a prediction for $\xi$ as a function of $1-E$. For $E$ we use the value measured on the largest lattice; the statistical errors and finite-size corrections on $E$ are less than $5 \times 10^{-5}$, and therefore induce a negligible error (less than 0.5%) on the predicted $\xi$. The corresponding observed/predicted ratios are also shown in Figure 2 (points □ and ◇). The "improved" 3-loop prediction is in excellent agreement with the data.

In Figure 3 we report the ratio

$$R_\chi(\beta) = \frac{C_\chi}{C^2_{\xi^{(2)}}} \frac{\chi_{\infty,estimate}/(\xi^{(2)}_{\infty,estimate})^2}{\chi_{\infty,theor}/(\xi^{(2)}_{\infty,theor})^2}, \quad (9)$$

where by the suffix *theor* we denote the one-, two- or three-loop prediction either for the standard perturbation theory in $1/\beta$ or for the "improved" one in $1-E$. The curves appear to become flat for increasing $\beta$ and to converge to the same value $\approx 10.8$, thus providing an estimate for the universal ratio $C_\chi/C^2_{\xi^{(2)}}$.

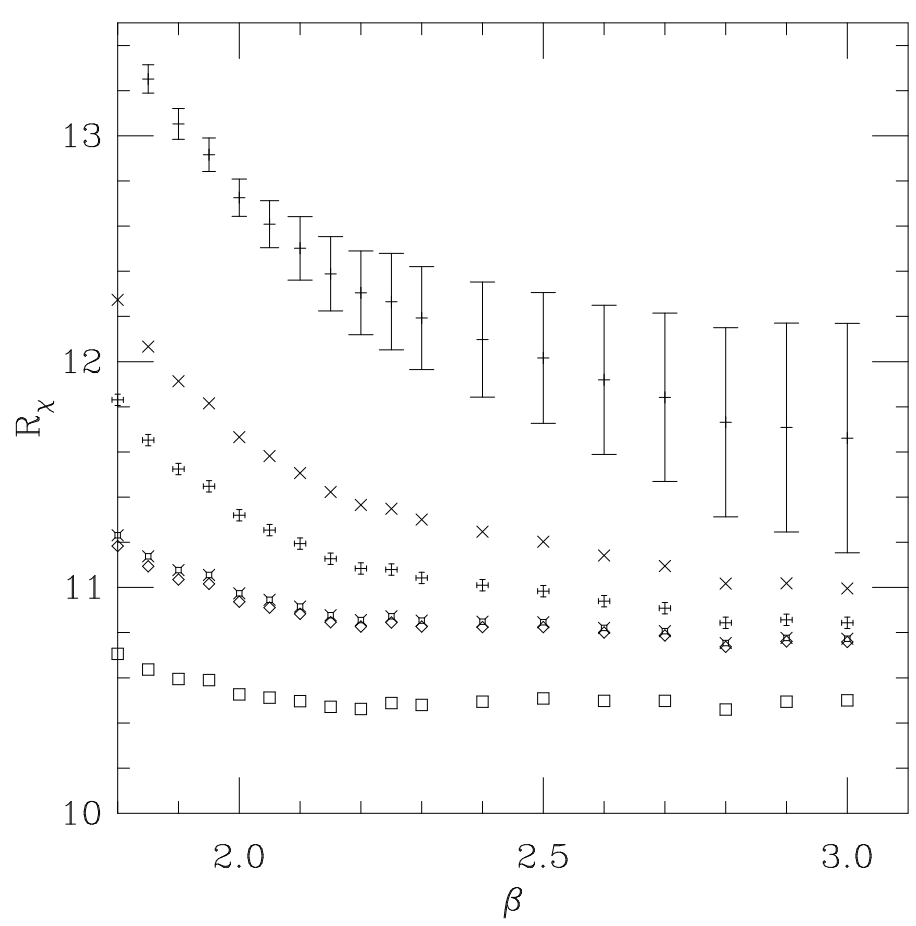

Figure 3. $R_\chi$ versus $\beta$. There are six versions of perturbation theory: the standard one in $1/\beta$ gives points $+$ (1-loop), $\times$ (2-loop) and ⊞ (3-loop); "improved" perturbation theory in $1-E$ gives points $\Box$ (1-loop), $\Diamond$ (2-loop) and ⧖ (3-loop). Error bars are one standard deviation and are reported only for one series of prediction because they are always the same at fixed $\beta$.

Let us summarize the conceptual basis of our analysis. The main assumption is that if the Ansatz (6) with a given function $F_\xi$ is well satisfied by our data for $L_{min} \leq L \leq 256$ and $1.65 \leq \beta \leq 3$, then it will continue to be well satisfied for $L > 256$ and for $\beta > 3$. Obviously this assumption could fail, e.g. if [14] at some large correlation length ($\gtrsim 10^3$) the model crosses over to a new universality class associated with a finite-$\beta$ critical point. In this respect our work is subject to the same caveats as any other Monte Carlo work on a finite lattice. However, it should be emphasized that our approach does *not* assume asymptotic scaling [eq. (1)], as $\beta$ plays no role in our extrapolation method. Thus, we can make an unbiased *test* of asymptotic scaling. The fact that we confirm (1) *with the correct nonperturbative constant* (3)/(4) is, we believe, good evidence in favor of the asymptotic-freedom picture.

We wish to thank Jae-Kwon Kim for sharing his data with us, and for challenging us to push to ever larger values of $\xi/L$. We also thank Steffen Meyer, Adrian Patrascioiu, Erhard Seiler and Ulli Wolff for helpful discussions. The authors' research was supported by CNR, INFN, DOE contracts DE-FG05-85ER250000 and DE-FG05-92ER40742, NSF grant DMS-9200719, and NATO CRG 910251.